\documentstyle[aps,psfig]{revtex}

\begin{document}

\title{Multibaryons in the Skyrme model\footnote{Cont. to the Proc. of "Hadron Physics 99".
Coimbra, Portugal. September 10- 15, 1999 to be published by AIP.} }

\author{Norberto N. Scoccola}

\address{
Physics Dept., Comisi\'on Nac. Energ{\'{\i}}a
At\'omica, Libertador 8250, (1429) Bs.As., Argentina\\
and\\
Universidad Favaloro, Sol{\'\i}s 453, (1078) Buenos Aires,
Argentina.
}

\maketitle

\begin{abstract}
Low-lying multibaryon configurations are studied within the bound state approach to the $SU(3)$ Skyrme model. We
use approximate ans\"atze for the static background fields based on rational maps which have the same symmetries of
the exact solutions. To determine the explicit form of the collective Hamiltonians and wave functions we only make
use of these symmetries. Thus, the expressions obtained are also valid in the exact case. On the other hand, the
meson bindings, inertia parameters and hyperfine splitting constants we calculate do depend on the detailed form of
the ans\"atze and are, therefore, approximate. Using these values we compute the low-lying spectra of multibaryons
with $B\le 9$ and strangeness $0$ and $-B$. With these results the stability of some multilambda configurations is
discussed.
\end{abstract}

\section{Introduction}

In the last few years there have been several important developments in the
determination of the lowest energy multiskyrmion
configurations\cite{KS87,BTC90,BS97}.
This type of solutions are essential for the understanding of multibaryons
and, perhaps, nuclei in the framework of the topological chiral soliton
models.
So far, these
models have proven to be useful for the description of quantities
such as the masses, strong and electromagnetic
properties of the octet and decuplet
baryons, baryon-baryon
interactions, etc. (see e.g. Refs.\cite{ZB86,Wei96} and references
therein). The
knowledge of the properties of the multiskyrmion configurations opens the
possibility
of studying more complex baryonic objects. In fact, several investigations
concerning non-strange multiskyrmion systems have been reported in the
literature
(see, e.g., Refs.\cite{BC86,Car91,Wal91,Wal96,Irw98}).
Of particular interest are, however, the strange multibaryons.
Perhaps the most celebrated example is the $H$ dibaryon
predicted in the context of the MIT bag model more than twenty years
ago\cite{Jaf77}.
This exotic has been studied in various other models, including the Skyrme model
\cite{BBLRS84,KSS92,KM88,TSW94}, but its existence remains controversial both
theoretically and experimentally.
It has also been speculated that strange matter could
be stable\cite{Wit84}. This has lead to numerous investigations of the
properties of
strange matter in bulk and in finite lumps (for a recent review see
Ref.\cite{GSB98}). Moreover,
with the new heavy ion colliders there is now the possibility of producing
strange multibaryons in the laboratory\cite{E864}. In this situation the
study of multibaryon systems within the $SU(3)$ Skyrme model appears to be
very interesting. For general soliton configurations
this is a quite hard numerical task since one has to deal with
several coupled partial differential equations. However, the problem is
greatly simplified if one introduces the (approximate) rational maps
ans\"atze\cite{HMS97}
for the multiskyrmion configurations. The construction of these ans\"atze is based
on the analogy between BPS monopoles and skyrmions and requires that the
approximate solutions have the same symmetries than the exact numerical
ones.
In fact, it is now known \cite{KS87,BTC90,BS97} that up to $B=9$ these configurations are
very symmetric. Namely, for $B=2$ the solution corresponds to an axially
symmetric torus while configurations with $B=3-9$ possess the symmetries
of the platonic polyhedra (e.g. tetrahedron for $B=3$, etc) . In contrast with the exact
solution, however, the rational map approximation assumes that the modulus
of the static pionic field is radially symmetric while its direction
depends only on the polar coordinates. In this contribution we
will report on how to describe multibaryon states in the $SU(3)$ Skyrme
model using these approximate ans\"atze.

\section{Symmetric multiskyrmions and rational maps}

A rational map of order $N$ is a map of $S^2 \to S^2$ of the form
\begin{eqnarray}
R_N(z) = \frac{p(z)}{q(z)}
\nonumber
\end{eqnarray}
where $p,q$ are polynomials of degree at most $N$ in the stereographic coordinate
$z = \tan(\theta/2) \ \exp( i \phi) $.
It was shown by Donaldson\cite{Don84} that there is a one-to-one correspondence between
BPS monopoles of order  $k$ and rational maps of degree $N=k$.
Using the analogy between this type of monopoles and the skyrmions,
the authors of Ref.\cite{HMS97} proposed the following ans\"atze for the static soliton chiral field
\begin{eqnarray}
U_N^{rat.} (\vec r) = \exp\left[ i \vec \tau \cdot \hat n_N \ F(r) \right]
\label{rat}
\end{eqnarray}
where
\begin{eqnarray}
\hat n_N = \left( \frac{2 \Re(R_N)}{1+|R_N|^2} ,  \frac{2 \Im(R_N)}{1+|R_N|^2}, \frac{1-|R_N|^2}{1+|R_N|^2} \right)
\end{eqnarray}

Replacing Eq.(\ref{rat}) in the Skyrme model effective action
\begin{eqnarray}
\Gamma_{eff} = \frac{f_\pi^2}{4} \int d^4x \ \mbox{Tr} \ \partial_\mu U \partial^\mu U^\dagger +
\frac{1}{32 e^2} \int d^4x \ \mbox{Tr} \left[ U^\dagger \partial_\mu U, U^\dagger \partial_\nu U \right]^2
\label{skyrme}
\end{eqnarray}
one gets the following expression for the soliton mass
\begin{eqnarray}
M_{sol} =
    {f_\pi^2\over{2}} \int d^3r  \left[ F'^2 + 2N\ {\sin^2 F \over{r^2}}
        \left( 1 + {F'^2\over{e^2 f_\pi^2}} \right) +
    { {\cal I}_N\over{e^2 f_\pi^2}}
        {\sin^4 F\over{r^2}} \right]
\end{eqnarray}
where
\begin{eqnarray}
{\cal I}_N \ = \
{1\over{4 \pi}} \int {2i\ dz d\bar z \over{ (1 + |z|^2)^2 }}
\ \left(
     {{1 + |z|^2} \over{ 1 + |R_N|^2 }}
          \left| {dR_N\over{dz}} \right| \right)^4
\end{eqnarray}

To obtain the ansatz for a given baryon number $B=N$ one should
proceed as follows. First, one constructs the most general map of degree $N$
that has the symmetries of the exact solutions. Then, the resulting ${\cal I}_N$
has to be minimized with respect to the remaining free
parameters. To perform the first step it is useful to recall that under a
general $SO(3)$ transformation the stereographic coordinate $z$ transforms as
\begin{eqnarray}
z \to \frac{\alpha \ z + \beta}{-\bar \beta \ z + \alpha}
\end{eqnarray}
where $\alpha, \beta$ are entries of the $J=1/2$ representation of the
corresponding rotation operator.
We illustrate the method by considering the case $B=2$. The most general map
of degree $N=2$ is
\begin{eqnarray}
R_2 = \frac{\mu \ z^2 + \nu \ z + \lambda}{\delta \ z^2 + \gamma \ z + \xi}
\end{eqnarray}
If we impose the symmetries of the exact torus configuration (axial symmetry
plus $\pi$ rotations around the three cartesian axes) such general
form reduces to
\begin{eqnarray}
R_2 = \frac{z^2 - a}{- a \ z^2 + 1}
\end{eqnarray}
The value of $a$ can be now determined by requiring that
it should minimize $M_{sol}$ (that is, ${\cal I}_2$). In this way one
finds $a=0$. Thus, the appropriate ansatz is
\begin{eqnarray}
R_2 = z^2
\end{eqnarray}
The explicit expressions of the rational maps corresponding to the other
baryon numbers have been given in Ref.\cite{HMS97}.
Once such maps are determined, the Euler-Lagrange
equation for the soliton profile $F(r)$ can be numerically
solved for each baryon number and the multiskyrmion masses $M_{sol}$
evaluated.

The values of the soliton masses (per baryon number) for the different baryon numbers as
calculated using the rational map ans\"atze are given in Table \ref{su2}. For reference, the
results corresponding to the skyrmion configurations which fully minimize
the static energies\cite{KS87,BTC90,BS97} and the associate symmetry groups are also given.
From this table one observes that the rational
map ans\"atze indeed provide a very good approximation to the
exact numerical solutions.

\section{Strange multibaryons}

\noindent
We turn now to the study of the strange multibaryons within the $SU(3)$ Skyrme model
using the rational map ans\"atze described in the previous section. For this purpose,
the effective action Eq.(\ref{skyrme})
has to be supplemented with the Wess-Zumino term and some suitable
flavor symmetry breaking terms. In the calculations described below we
have included terms that account for the different pseudoscalar meson masses and
also for the difference between their decay constants.

To extend the model to $SU(3)$ flavor space we use the bound state approach, in which
strange baryons appear as bound kaon-soliton systems\cite{CK85}.
Thus, we introduce a generalized Callan-Klebanov ansatz
\begin{eqnarray}
U=\sqrt{U_{N}}\ U_K \ \sqrt{U_{N}}
\label{ans}
\end{eqnarray}
where $U_{N}$ is the $SU(2)$ multiskyrmion field properly embedded
into $SU(3)$ and $U_K$ is the field that carries the strangeness. Its form is
\begin{eqnarray}
U_K \ = \ \exp \left[ i\frac{\sqrt2}{f_K} \left( \begin{array}{cc}
                            0 & K \\
                            K^\dag & 0
                           \end{array}
                       \right) \right]
\end{eqnarray}
where $K$ is the usual kaon isodoublet.

In the spirit of the bound state approach we consider first the problem of a kaon
field in the background of a static multiskyrmion configuration. To describe such
configuration we use the rational map ansatz approximation Eq.(\ref{rat}).
Consequently, the ansatz for the kaon field should be
\begin{eqnarray}
K = k_N(r,t) \  \vec \tau \cdot \hat n \ \chi
\label{kaon}
\end{eqnarray}
where $\chi$ is a 1/2 spinor. Replacing Eqs.(\ref{ans}-\ref{kaon})
in the effective action and performing
the corresponding canonical transformations we obtain a quadratic Hamiltonian whose
diagonalization leads to\cite{SS98}
\begin{eqnarray}
\left[ - {1\over{r^2}} \partial_r \left( r^2 h \partial_r \right)
+ m_K^2 + V - f \epsilon_N^2  - 2 \ \lambda \ \epsilon_N
\right] k(r) = 0
\label{eigen}
\end{eqnarray}
The radial functions $f$, $h$, $\lambda$ and $V$ depend on the baryon number
$B$ only through the integral ${\cal I}_N$. Their explicit
expressions can be found in Ref.\cite{SS98}.

Eq.(\ref{eigen}) has been solved numerically for different values of
$B$ using the values of Ref.\cite{ANW83} for $f_\pi$ and $e$ and setting
$m_K$ and $f_K/f_\pi$ to their corresponding empirical values.
The resulting eigenenergies are listed in Table \ref{ta2}.
Also listed are the masses (per baryon number) of the corresponding
$Y=0$ states in the adiabatic approximation,
$M_{Y=0}^{adiab}/B = M_{sol} + \epsilon$. These states
are of particular interest since it has been
claimed\cite{Jaf77,IKS88} that some of them can be stable against
strong decays. As a general trend we see that the kaon binding energies
$D^K_N = m_K - \epsilon_N$ decrease with increasing baryon number.
However, as in the case of the energy required to liberate a single $B=1$ skyrmion
from the multisoliton background\cite{BTC90,BS97},
we observe some deviation from a smooth behaviour,
namely, $D^K_4 > D^K_3$ and $D^K_7 > D^K_6$.
Consequently, such deviations will be also present in the multiskyrmion
mass per baryon. Interestingly, this kind of phenomena has been also observed
in some MIT bag model calculations\cite{GSB98}. There they are
due to shell effects.

Using the values given in Table \ref{ta2} we obtain
\begin{eqnarray}
 \begin{array}{ccc}
M_{2\Lambda} - 2 M_{\Lambda} &=& 12 \ MeV \\
M_{4\Lambda} - 2 M_{2\Lambda}  &=& -176 \ MeV \\
M_{7\Lambda} - ( M_{3\Lambda} - M_{4\Lambda} ) &=& - 177 \ MeV
\end{array}
\end{eqnarray}
in the static soliton approximation (i.e. to ${\cal O}( N_c^0)$).
These results seem to confirm previous speculations about the
stability of the tetralambda in the Skyrme model\cite{IKS88}
and opens up the possibility of a stable heptalambda.
On the other hand, they indicate that the
$H$-particle, although very close to threshold, is not stable.

Within the static multiskyrmion approximation considered so far
the spin and isospin quantum numbers of the bound kaon-multiskyrmion systems
are not well defined.
To recover good spin and isospin quantum numbers we proceed with
the standard semi-classical collective quantization\cite{ANW83}.
For $B >1$, however, we should introduce independent spin and
isospin rotations. The collective Lagrangian reads
\begin{eqnarray}
 \begin{array}{ccc}
    L_{coll} & = & {1\over2} \left[ \Theta^J_{ab} \ \Omega_a \Omega_b +
            \Theta^I_{ab} \ \omega_a \omega_b +
        2 \ \Theta^M_{ab} \ \Omega_a \omega_b \right]
         - \left( c^J_{ab} \Omega_a + c^I_{ab} \omega_a \right) \ T_b
                           \end{array}
\label{lcol}
\end{eqnarray}
Here, $\vec \Omega$ is the angular velocity corresponding to the spin rotation,
$\vec \omega$ that of the isospin rotation and $T_b$ is the kaon spin.
$\Theta^J_{ab}$ and $\Theta^I_{ab}$ are the corresponding
moments of inertia while $\Theta^M_{ab}$ is an inertia that mixes spin and
isospin. The constants $c^J_{ab}$ and $c^I_{ab}$ are the  hyperfine
splitting constants which for $B=1$ provide the $\Lambda$-$\Sigma$ mass splitting.
The explicit expressions of these inertia and hyperfine splitting tensors
in terms of the soliton profile function $F(r)$ and the rational
map $R_N(z)$ can be found in Ref.\cite{GSS99}.

Using the standard definitions for the canonical conjugate momenta
\begin{eqnarray}
 \begin{array}{ccc}
J_a &=& \frac{\partial L_{coll}}{\partial \Omega_a}
= \Theta_{ab}^J \ \Omega_b + \Theta_{ab}^M \ \omega_b - c_{ab}^J \ T_b  \\[.5cm]
I_a &=& \frac{\partial L_{coll}}{\partial \omega_a}
= \Theta_{ab}^M \ \Omega_b + \Theta_{ab}^I \ \omega_b - c_{ab}^I \ T_b
                           \end{array}
\end{eqnarray}
it is rather simply to find the general form of the collective
Hamiltonian $H_{coll}$. Details are given in Ref.\cite{GSS99}.

It is important to stress that the structure of the inertia and hyperfine splitting
tensors appearing in Eq.(\ref{lcol}) is strongly determined by the multiskyrmion symmetries.
Using group theory arguments, it can be shown that
(for symmetric skyrmions) such tensors are always diagonal. The number
of independent diagonal entries, as well as whether the mixing
inertias vanish or not, is also fixed by the properties of the corresponding symmetry group $G$.
For example, for $B=3$ the three components of the spin and isospin operators
transform as the 3-dim irrep $F_2$ of the group $T_d$. Therefore, there is only one
independent component for the spin inertia, one for isospin inertia and one
for the mixing inertia. Similar analysis can be done for the hyperfine splittings.
For $B=4$, however, $I_1, I_2$ transform as the 2-dim irrep $E_g$ of the group $O_h$
while $I_3$ as the 1-dim irrep $A_{2g}$ and the three components of $\vec J$
as the 3-dim irrep $T_{1g}$. Thus, for $B=4$ we should have
\begin{eqnarray}
\Theta^I_{11} = \Theta^I_{22} \ne \Theta^I_{33} \ ;
\qquad \Theta^J_{11} = \Theta^J_{22} = \Theta^J_{33} \ ;
\qquad \Theta^M_{aa} = 0
\end{eqnarray}

Finally, we have to determine the collective wave-functions. Their
general form must be
\begin{eqnarray}
| J J_z, I I_z, S \rangle = \sum_{J_3 I_3 T_3} \,
    \beta^{JIT}_{J_3 I_3 T_3} \,
    D^J_{J_z J_3} \, D^I_{I_z I_3} \, K^T_{T_3}
\nonumber
\end{eqnarray}
where $ D^J_{J_z J_3}$ and $D^I_{I_z I_3}$ are $SU(2)$ Wigner functions
and  $\beta^{JIT}_{J_3 I_3 T_3}$ are some numerical coefficients that
have to be fixed by requiring that these wave-functions
transform as a 1-dim irrep of $G$. It is very important
to notice that such irrep may not coincide with the trivial
irrep. As well known when one performs an
adiabatic symmetry operation on a skyrmion configuration one can pick
a non-trivial phase. These are the so-called Filkenstein-Rubinstein
phases. A detailed analysis of these phases for the configurations
we are dealing with has been done by Irwin\cite{Irw98}.
Using these phases one gets that, except for $B=5,6$, the wavefunctions should
transform as the trivial irrep of $G$. For $B=5$ they should transform
as the $A_2$ irrep of $D_{2d}$ and for $B=6$ as the $A_2$ irrep of $D_{4d}$.

Having obtained the explicit form of the collective Hamiltonians and wavefunctions,
the ${\cal O}( N_c^{-1})$ rotational contribution  $E_{rot}$ to the
multiskyrmion masses can be calculated using first order perturbation theory.
The numerical values of such contributions to the masses of the
lowest lying non-strange baryons are given
in Table \ref{nonstr}
while those corresponding to the zero-hypercharge multibaryons are listed
in Table \ref{ta4}.

From Table \ref{nonstr} we note that the quantum numbers of the ground states are
consistent with those known for light nuclei with the exception of the
odd values $B=5,7,9$. We also observe that the lowest lying state has the lowest
possible value of isospin and on the average mass splittings decrease for
increasing baryon number. This is a consequence of the fact that, although
all the moments of inertia increase with increasing baryon number,
the increase of the spin inertia is much faster than that of
the isospin one.

Using the values of the rotational corrections to the lowest lying $Y=I=0$
states (some of which are listed in Table \ref{ta4}) one can see that
stability of the $4\Lambda$ and the $7\Lambda$ is not affected by these
corrections. For example, for the $4\Lambda$ there is a decrease of
$36 \ MeV$ in the binding energy while that of the $7\Lambda$ is increased
by $45 \ MeV$.

\section{Conclusions}

In this contribution we have reported on the description of multibaryons
within the bound state approach to the $SU(3)$ Skyrme model. To describe the multiskyrmion
backgrounds we have used ans\"atze based on rational maps. Such configurations are known
to provide a good approximation to the exact numerical ones, and lead to
a great simplification in the treatment of the kaon-soliton system. An important
property of these approximate configurations is that they have the same symmetries
as the exact ones. We have shown that the properties of the associated symmetry groups
completely determine the explicit form of the collective Hamiltonians
(namely, the detailed structure of the inertia and hyperfine splitting tensors).
The same happens for the collective wavefunctions. In particular, we have shown how the
Filkenstein-Rubinstein phases fix, in a unique way, the one dimensional irreducible
representations as which each wave function should transform.
Thus, the method to obtain the collective Hamiltonians and wave functions described
here is also valid in the exact case. On the other hand, the numerical values of
the meson bindings and of the independent inertia parameters and hyperfine splitting
constants will depend on the detailed form of the ans\"atze and will be,
therefore, approximate.

Using an effective action that provides a good description of the hyperon
static properties we have studied the spectra of non-strange and strange
multibaryons. In the case of non-strange baryons we found that,
for even baryon number, the  ground state quantum numbers coincide with
those of known stable nuclei. It should be stressed, however, that
in our opinion these quite compact multiskyrmion configurations should
be interpreted as "multiquark bags"
rather than normal nuclei. How these configurations are related with
them is not yet clear.  Another feature of the predicted spectra is that the low
lying non-strange multibaryons always have the lowest possible value of isospin.
This can be understood in terms of the behaviour of the inertia tensors
as a function of the baryon number. The situation is more complicated in the
case of strange particles for which there is a quite delicate interplay between the
different terms contributing to the rotational energies.
From the calculated spectra of strange multibaryon it results that some $Y=0$
configurations could be stable against strong decays. Such configurations,
usually called strangelets, are expected to be seen in RHIC\cite{GSB98,E864}.

Many of the ideas discussed in the present
contribution can be extended to case of heavy flavor (e.g. charmed)
multibaryons\cite{KW99}. In such case, however, a proper treatment requires
the use of an effective Lagrangian
that accounts for both chiral symmetry and heavy quark symmetry.
The present model has been also applied to the study of the binding
of the $\eta$ meson to few non-strange baryon systems\cite{RS98}.

We finish with a comment on the Casimir corrections to the multibaryon masses. Although
these corrections are not expected to affect in any significant way the
kaon eigenvalues and the rotational energies shown here, they might play some role in the determination of the
multibaryon binding energies. Within the $SU(2)$ Skyrme model it has been shown\cite{MK91} that
they are responsible for the reduction of the otherwise large $B=1$ soliton mass to a reasonable value
when the empirical value of $f_\pi$ is used. Here, we have avoided the $B=1$ large mass problem by using
the customary method of fitting $f_\pi$ to reproduce the nucleon mass\cite{ANW83}.
A more consistent approach should certainly use the empirical $f_\pi$ and include
the Casimir corrections. In this respect, there have been recently some
efforts\cite{Wal98} to evaluate the corrections to the $B=1$ mass in the $SU(3)$ Skyrme
model. Unfortunately, even in the $SU(2)$ sector, almost nothing is known for $B > 1$.
This is, of course, a very difficult task since it requires the
knowledge of the meson excitation spectrum around the non-trivial multiskyrmion up to rather
large energies.

\section*{ACKNOWLEDGEMENTS}

The material presented here is based on work done with J.P. Garrahan and
M. Schvellinger.  Support provided by the grant
PICT 03-00000-00133 from ANPCYT, Argentina is acknowledged. The author is
fellow of the CONICET, Argentina. He would like to thank the members
of the Organizing Committee for their warm hospitality during the workshop.

\begin{table}

\caption{Soliton mass per baryon number (in natural units $= 6 \pi^2 f_\pi/e$) obtained by using the rational map
ansatz (APPROX) as compared with the (EXACT) numerical minimization. The corresponding symmetry group $G$ is also
listed.}

\label{su2}

\begin{center}
\begin{tabular}{ccccc}
\hspace{.5cm} $B$ \hspace{.5cm} & \hspace{.5cm}$\cal I$ \hspace{.5cm} & \hspace{.5cm} APPROX \hspace{.5cm} &
\hspace{.5cm} EXACT \hspace{.5cm} & \hspace{.5cm} $G$  \hspace{.5cm} \\ \tableline
 $1$ & $1$      & $1.232$ & $1.232$ & $O(3)$ \\
 $2$ & $5.81$   & $1.208$ & $1.171$ & $D_{\infty,h}$ \\
 $3$ & $13.58$  & $1.184$ & $1.143$ & $T_d$ \\
 $4$ & $20.65$  & $1.137$ & $1.116$ & $O_h$ \\
 $5$ & $35.75$  & $1.147$ & $1.116$ & $D_{2d}$ \\
 $6$ & $50.76$  & $1.137$ & $1.109$ & $D_{4d}$ \\
 $7$ & $60.87$  & $1.107$ & $1.099$ & $Y_h$ \\
 $8$ & $85.63$  & $1.118$ & $1.100$ & $D_{6d}$ \\
 $9$ & $112.83$ & $1.123$ & $1.099$ & $T_d$ \\
\end{tabular}
\end{center}

\end{table}

\begin{table}

\caption{Kaon eigenenergy $\epsilon_N$ and adiabatic approximation to the mass of the zero-hypercharge states (all
in $MeV$) as a function of the baryon number $B$.}

\label{ta2}

\begin{center}
\begin{tabular}{ccc}
\hspace{1cm} $B$ \hspace{1cm} & \hspace{1cm} $\epsilon_N$  \hspace{1cm} & $M_{Y=0}^{adiab.}/B$ \\  \tableline
 $1$ & $222$ & $1085$\\
 $2$ & $244$ & $1091$ \\
 $3$ & $255$ & $1085$ \\
 $4$ & $250$ & $1047$ \\
 $5$ & $263$ & $1067$ \\
 $6$ & $267$ & $1064$ \\
 $7$ & $262$ & $1038$ \\
 $8$ & $271$ & $1055$ \\
 $9$ & $276$ & $1063$ \\
\end{tabular}
\end{center}

\end{table}

\begin{table}

\caption{Quantum numbers and rotational energies of the lowest lying $S=0$ states.}

\begin{center}
\begin{tabular}{cccc}
\hspace{.9cm} $B$ \hspace{.7cm} & \hspace{.7cm} $J^P$  \hspace{.7cm} & \hspace{.7cm} $I$ \hspace{.7cm} &
\hspace{.7cm} $ E_{rot} [MeV]$ \hspace{.4cm} \\ \tableline
 3     &${1/2}^+$ & ${1/2}$ &  64   \\
       &${5/2}^-$ & ${1/2}$ & 147  \\[2.mm]
 4     &$ 0^+   $ &    0    &   0   \\
       &$ 4^+   $ &    0    & 173  \\[2.mm]
 5     &${1/2}^+$ & ${1/2}$ &  28  \\
       &${3/2}^+$ & ${1/2}$ &  40  \\[2.mm]
 6     &$ 1^+   $ &    0    &   7  \\
       &$ 3^+   $ &    0    &  44  \\[2.mm]
 7     &${7/2}^+$ &${1/2}$&  66 \\
       &${3/2}^+$ &${3/2}$&  98  \\[2.mm]
 8     &$  0^+  $ &   0   &   0 \\
       &$  2^+  $ &   0   &  14 \\[2.mm]
 9     &$  1/2^+    $ &   1/2   &   14  \\
       &$  5/2^-    $ &   1/2   &   30 \\[2.mm]
\end{tabular}
\end{center}
\label{nonstr}

\end{table}

\begin{table}

\caption{Quantum numbers and rotational energies of the lowest lying $Y=0$ states.}

\label{ta4}

\begin{center}
\begin{tabular}{cccc}
\hspace{.9cm} $B$ \hspace{.7cm} & \hspace{.7cm} $J^P$  \hspace{.7cm} & \hspace{.7cm} $I$ \hspace{.7cm} &
\hspace{.7cm} $ E_{rot} [MeV]$ \hspace{.4cm} \\ \tableline
 3     &${1/2}^+$ &    1    &  50   \\
       &${3/2}^-$ &    0    &  77  \\[2.mm]
 4     &$ 0^+   $ &    2    &   51  \\
       &$ 0^+   $ &    0    &   72 \\[2.mm]
 5     &${1/2}^+$ &    1    &  29 \\
       &${1/2}^-$ &    1    &  32 \\[2.mm]
 6     &$ 0^+   $ &    2    &   24 \\
       &$ 0^-   $ &    1    &   26 \\[2.mm]
 7     &${3/2}^+$ &   2   &  32 \\
       &${5/2}^+$ &   1   &  65 \\[2.mm]
 8     &$  0^+  $ &   2   &   19 \\
       &$  2^+  $ &   2   &   31 \\[2.mm]
 9     &$  1/2^-    $ &   2     &   25 \\
       &$  3/2^-    $ &   2     &   29 \\
\end{tabular}
\end{center}

\end{table}

\end{document}